\documentclass{ws-mpla}
\usepackage[super]{cite}
\usepackage{graphicx}
\usepackage{float}
\usepackage{color}
\usepackage{xcolor}
\usepackage{longtable}
\usepackage[colorlinks,citecolor=blue,linkcolor=red,anchorcolor=blue,filecolor=blue,urlcolor=blue]{hyperref}
\usepackage{graphicx}
\usepackage{dcolumn}
\usepackage{bm}
\usepackage{hyperref}
\usepackage[mathlines]{lineno}

\begin{document}
\markboth{Nishu Jain, M. Bhuyan, and Raj Kumar}{Exploring the role of low-lying intrinsic degrees of freedom and their impact on fusion cross-sections
}
\catchline{}{}{}{}{}

\title{Exploring the role of low-lying intrinsic degrees of freedom and their impact on fusion cross-sections }
\author{Nishu Jain$^{1}$, M. Bhuyan$^{2}$, Raj Kumar$^1$}
\address{$^1$Department of Physics and Material Science, Thapar Institute of Engineering and Technology, Patiala 147004, India  \\}
\address{$^2$Institute of Physics, Sachivalya Marg, Bhubaneswar-751005, Odisha, India}
\date{\today}
\maketitle
            
\begin{abstract}
\noindent
The present work focuses on examining the low-lying intrinsic degrees of freedom and their impact on fusion dynamics. Fusion cross-sections were calculated using the coupled-channel code CCFULL for four specific reactions: $^{18}$O+$^{74}$Ge, $^{18}$O+$^{148}$Nd, $^{18}$O+$^{182}$W, and $^{18}$O+$^{186}$W, all conducted at energies below the Coulomb barrier across various energy levels. Vibrational and rotational features were studied concerning energy to distinguish their respective effects on fusion properties. The results indicate that the theoretical calculations for the nuclei $^{74}$Ge, $^{148}$Nd, $^{182}$W and $^{186}$W closely match the experimental data, particularly for the $2^+$ excited states. While slight discrepancies are observed for other excited states ($4^+$ and $6^+$), overall agreement remains significant. Additionally, the study reveals that hexadecapole deformation with different magnitudes have significant influences on the fusion cross-section. In cases where $\beta_4$ has a positive value, rotational levels beyond $6^+$ have minimal impact on the cross-section, resulting in a notable difference in the contribution of sequential channels. In contrast, for negative $\beta_4$ values, rotational energy levels up to the $2^+$ state substantially affect the fusion characteristics. Furthermore, the analysis extends to the estimation of the relative change ($\Delta\sigma_{fus}$) between the excited states and the ground state, both with and without considering coupling terms.
\end{abstract}
\noindent
\noindent

\section{Introduction}
\label{intro} \noindent 
Recent advances in heavy-ion fusion reactions have paved the way for an exciting avenue of scientific exploration, providing valuable insights into the characteristics and behaviours of the heaviest elements \cite{hind21}. A thorough examination of nuclear fusion processes is imperative in our ongoing pursuit of unravelling the mysteries of stellar evolution and understanding the mechanisms underlying the colossal power generated by stars. The fusion dynamics at subbarrier energies are notable for being shaped by a myriad of intrinsic degrees of freedom, including deformation, neutron transfer channels, neck formation, and vibrations, underscoring the intricate nature of the fusion process \cite{rajb16, rowl92, stef95a, cham02, saha96}. Extensive investigations by various researchers have illustrated the pivotal role played by these intrinsic degrees of freedom in enhancing the sub-barrier fusion cross-section \cite{beck98,dasg98,bala98,mont13,back14,cant20}. Experimental investigations of fusion excitation functions across various combinations of projectiles have revealed a significant discrepancy with the one-dimensional barrier penetration model (1-D BPM), indicating a notable underprediction of fusion rates. This underscores the limitations of the 1-D BPM in accurately capturing sub-barrier fusion dynamics \cite{dasg98,kola12, mont17}. Recent studies suggest that the primary cause of this disparity stems from the interplay among the intrinsic degrees of freedom of the reaction partners, leading to a modification of the nominal fusion barrier into a distribution of barriers with an average height. This distribution termed the fusion barrier distribution, introduces lower barriers that offer a more favourable pathway for the incoming flux to access the fusion channel, thereby resulting in an amplified fusion cross-section. This underscores the pivotal role of the fusion barrier distribution in influencing sub-barrier fusion dynamics.

Harnessing its capability to scrutinize the impact of specific degrees of freedom on the fusion process, the Coupled Channels (CC) approach stands out as a significant enhancement in fusion studies, particularly in the sub-barrier region. This methodology proves invaluable for delving into the coupling mechanisms responsible for heightened fusion cross-sections at energies below the Coulomb barrier, as it provides insights into the fusion barrier distribution. However, attempts to establish a direct correlation between intrinsic degrees of freedom and the sequence of fusion enhancement have yielded inconclusive results. This underscores the necessity for further exploration employing analytical and empirical approaches to gain a deeper understanding of this phenomenon. The selection of nuclear potential emerges as a pivotal aspect in theoretical models, exerting a substantial influence on the nominal barrier width and tunnelling phenomenon, thus favouring the fusion process \cite{fern89, dasg92, hagi99}. In the existing literature, various forms of the nucleus-nucleus potential \cite{cham02, saha96, bloc77, aver18, akyu81, duha11} have been used to investigate heavy-ion fusion dynamics, including the utilization of the Woods-Saxon (WS) potential. The WS potential, defined by three adjustable parameters—range ($r_0$), depth ($V_0$), and diffuseness ($a_0$)—is a prominent choice in these studies.
\begin{equation}
    V_{N}= \frac{-V_0}{1+\exp\Big[\big(r_0-R_0\big)/a_0\Big]},  \\ 
    \label{ws}
\end{equation}
The parameters of the Woods-Saxon (WS) potential are determined by fitting the theoretical outcomes to the experimental data, particularly focusing on energies above the barrier. Additionally, this investigation will explore the influence of low-lying intrinsic degrees of freedom and their structural aspects on fusion dynamics, particularly at energies below the Coulomb barrier, considering both spherical and deformed nuclei. Fusion cross-sections were computed using the Coupled Channels code CCFULL \cite{hagi99} for four distinct reactions: $^{18}$O+$^{74}$Ge, $^{18}$O+$^{148}$Nd,  $^{18}$O+$^{182}$W, and $^{18}$O+$^{186}$W, all conducted at energies below the Coulomb barrier at various energy levels. The analysis incorporates vibrational and rotational characteristics in terms of energy to ascertain their respective impacts on fusion properties. It is important to note that the current calculations do not include nucleon transfer channels, which could be significant at energies below the Coulomb barrier \cite{jain2024,trip01} and associated references. Furthermore, a comparative assessment will be made between the resulting theoretical predictions and the available experimental data for the considered nuclear reactions.
\section{Coupled channel Approach}
\label{theory} \noindent 
The coupled-channel formalism, implemented in the CCFULL code, offers a robust framework for adequately describing the fusion excitation function in specific reactions under scrutiny. To streamline the coupled-channel equations, the no-Coriolis approximation is frequently employed, reducing the dimensionality of the equations. This involves substituting the total angular momentum (J) for the angular momentum associated with each channel's relative motion, facilitating a more straightforward solution of the channel equations,
\begin{eqnarray}
\Bigg[\frac{-\hbar^2}{2\mu} \frac{d^2 }{dr^2} &+& \frac{J(J+1)\hbar^2}{2\mu r^2}+\frac{Z_P Z_T e^2}{r}+V_{N}+\epsilon_n -E_{c.m.}\Bigg] \nonumber \\
&& \times \psi_n (r) + \sum_m V_{nm} (r)\psi_m (r) =0. 
\label{cc1}   
\end{eqnarray}
Eqn. (\ref{cc1}) encompasses several parameters essential to the coupled-channel (CC) equation, including the reduced mass ($\mu$) of the colliding nuclei, the radial component of the relative motion ($r$), the excitation energy of the $n^{th}$ channel ($\epsilon_n$), the coupling matrix Hamiltonian ($V_{nm}$), the nuclear potential ($V_N$), $E$ is the bombarding energy in the center of mass frame, $Z_P$ and $Z_T$ are the atomic number for projectile and target, respectively. And $k_0$ = $\frac {\sqrt{2\mu E}}{\hbar}$ is the wave number associated with the energy $E$. The computations are conducted using the CCFULL code, which accounts for the internal degrees of freedom associated with the structural characteristics of the reaction partners through the incorporation of a Woods-Saxon potential, as previously mentioned. By applying incoming wave boundary conditions (IWBCs) and assuming no-Coriolis effects, the penetrability of the barrier for each partial wave is calculated, and the resulting fusion cross-sections are expressed, incorporating all nonlinear coupling terms, as elaborated in Ref. \cite{hagi99}.
\begin{eqnarray}
\sum \sigma_J (E)=\sigma_{fus} (E)=\frac{\pi}{k_{0}^{2}}\sum_{J}(2J+1)P_J (E).
\label{fusion}
\end{eqnarray}
Here, the transmission coefficient for $J$ is represented by $P_J(E)$.
\section{Calculation and Discussions}
\label{result}\noindent
Our investigation entails conducting coupled-channel calculations that incorporate various degrees of freedom, encompassing vibrational and rotational aspects, for four distinct reactions: $^{18}$O+$^{74}$Ge, $^{18}$O+$^{148}$Nd, $^{18}$O+$^{182}$W, and  $^{18}$O+$^{186}$W. The primary objective of this study is to establish a correlation between the dynamics of reactions and the structural attributes of different nuclei. The analytical process begins with a thorough examination of the structure of both the target and the projectile. In this study, the projectile utilized is $^{18}$O, characterized as a vibrational nucleus with an excitation energy of 1.982 MeV for its first $2^+$ excited state and a quadrupole deformation ($\beta_2$) of 0.355 \cite{rama01}. The target nuclei under consideration, namely $^{74}$Ge, $^{148}$Nd, $^{182}$W, and $^{186}$W, exhibit both vibrational and rotational characteristics. Given the presence of rotational bands in their ground state, they are classified as rotational nuclei. 

\begin{table}
\centering
\caption{\label{table1}{Theoretical and experimental excitation energies corresponding to different entrance channels for $^{74}$Ge, $^{148}$Nd, $^{182}$W, and $^{186}$W}}
\renewcommand{\tabcolsep}{0.1cm}
\renewcommand{\arraystretch}{1.0}
\begin{tabular}{cccccc}
\hline \hline 
Nucleus & $E_2^+$ & Channel & \multicolumn{3}{c}{Excitation Energy (E)} \\
&  &  ($I$) & $E_{theo.}$ & $E_{exp.}$ & $\Delta E$ \\
(MeV) & MeV &  & (MeV) & (MeV) & ($E_{exp.}$ - $E_{theo.}$) \\
\hline
$^{74}$Ge & 0.59 & $2^+$ & 0.60 & 0.595  & 0.00\\
& & $4^+$ & 1.98 & 1.463  & 0.52\\
& & $6^+$ & 4.17 & 2.569 & 1.60\\
\cline{1-6}
$^{148}$Nd & 0.30 & $2^+$ & 0.30 & 0.30  & 0.00\\
& & $4^+$ & 1.00 & 0.45  & 0.55\\
& & $6^+$ & 2.11 & 0.53  & 1.58\\
\cline{1-6}
$^{182}$W & 0.10 & $2^+$ & 0.10 & 0.10  & 0.00\\
& & $4^+$ & 0.33 & 0.329  & 0.00\\
& & $6^+$ & 0.70 & 0.68 & 0.02\\
\cline{1-6}
$^{186}$W & 0.12 & $2^+$ & 0.12 & 0.12  & 0.00\\
& & $4^+$ & 0.41 & 0.40  & 0.01\\
& & $6^+$ & 0.81 & 0.85 & 0.04\\
\hline \hline 
\end{tabular}
\label{table1}
\end{table}

To enhance our understanding of the nuclear structure of the target nuclei, we conducted a comparative analysis between the theoretical and experimental excitation energies across various rotational energy levels. These energies are calculated using the rigid rotor expression, as outlined in Table \ref{table1}. The formula for the rigid rotor $E(I) = \frac{1}{6}E(2_1^+)I(I+1)$ is provided by \cite{Kran86}. Note: This formula is specifically used for the rigid rotor (nuclei which undergo rotational energy levels but not vibrational energy levels). Furthermore, to depict the energy levels of the nuclei considered, we utilize the notation $I$ for the entrance channel, and $E(2_1^+)$ denotes the energy of the first excited state. Furthermore, the energy level associated with the $I^+$ state can be expressed as $I(I+1)$ in terms of excitation energy. For example, in the context of even-even nuclei such as $^{74}$Ge, and $^{148}$Nd the calculated and experimental low-lying energy levels for $2^+$, $4^+$, and $6^+$ states are detailed in Table \ref{table1}. Theoretical calculations for $^{74}$Ge and $^{148}$Nd demonstrate satisfactory agreement with the experimental findings, especially for the $2^+$ excited state. However, notable relative changes are observed for the positive parity state $4^+$, with percentages of 26.6\% and 55\% for $^{74}$Ge and $^{148}$Nd nuclei, respectively. Furthermore, for the positive parity state $6^+$, this change exceeds around 75\% for both nuclei. This departure from rigid-rotor expression suggests that these nuclei are more aptly characterized as vibrational or approaching a rotational nature. According to Ref. \cite{ibbo97}, $^{148}$Nd is identified as a shape-transitional nucleus, while $^{74}$Ge is the crucial element that signifies the transition from a flexible to a rigid structure in Ge isotopes. The characteristics of individual levels are expounded based on the vibrational and rotational models. The intricate triaxial evolution with spin in $^{74}$Ge unfolds through systematic comparisons and analysis, identifying it as the pivotal nucleus heralding the transition from soft to rigid in Ge isotopes \cite{sun14}.

Incorporating two additional nuclei, $^{182}$W and $^{186}$W reveals that theoretical calculations for $^{182}$W align well with the experimental data, particularly for the excitation energy of the specific $2^+$ state. However, a discrepancy of 14.2\% and 24.3\% exists between the experimental and theoretical values for the $2^{nd}$ and $3^{rd}$ excitation states. The experimental excitation energies are sourced from Ref. \cite{nndc}. Similarly, for $^{186}$W, the theoretically computed excitation energy for the $2^+$ state closely matches the experimental value. In addition, there is a negligible deviation for the individual $4^+$ and $6^+$ rotational states, with relative changes of almost $(2 \%)$, respectively. Consequently, when considering the influence of individual rotational energy levels, it is crucial to select an appropriate set of nuclei whose rotational structure closely approximates that of a pure rotor. The study of nuclear shape phases is an essential tool for understanding the colliding motion of nuclei, and the behaviour of the energy ratios between different states along an isotopic chain provides a valuable signature of shape transitions \cite{khal13}. By employing the appropriate formula, it is possible to calculate the shape transition between $2^+$ and $4^+$ excitation states,
\begin{equation}
    R(4/2) = \frac{E(4_1^+)}{E(2_1^+)}.
    \label{ratio}
\end{equation}

The energy ratio $R(4/2)$ proves to be a valuable indicator for discerning shape transitions, ranging from a value of 2 for vibrational nuclides with spherical shapes to a characteristic value of 3.33 for well-deformed rotors. In the rapid transitional region, there is a distinct shift in the energy ratio $R(4/2)$ (equals 2). Experimental observations of target nuclei, including $^{186}$W, $^{182}$W, $^{148}$Nd, and $^{74}$Ge, reveal $R(4/2)$ ratios of 3.25, 3.29, 1.50, and 2.46, respectively. According to Ref. \cite{khal13}, $^{182}$W and $^{186}$W, with $R(4/2)$ = 3.25, and 3.29 falls under the category of rotational nuclei. However, $^{74}$Ge and $^{148}$Nd exhibit $R(4/2)$ ratios of 2.46 and 1.50, respectively, situating $^{148}$Nd in the intermediate range between magic nuclei; however, $\gamma$ instability shows in the case of the $^{74}$Ge nuclei. The disparity between experimental and theoretical results suggests a vibrational nature in these instances. Despite the proximity of the excitation energies of $^{74}$Ge and $^{148}$Nd to those of vibrational nuclei, they also feature $4^+$ and $6^+$ states, with excitation energies of at least three individual rotational levels closely aligning with experimental values. Consequently, these states are considered rotational, up to $3$ channels, to discern the impact of each energy level. Our analysis underscores that reliable classification as rotational occurs when there is substantial concordance between calculated and experimental excitation energies. Delving into the higher levels enumerated in the NNDC \cite{nndc} can offer deeper insights into the structure, building upon the information available in the lower three bands.
\begin{table} 
\centering
\caption{\label{table2} The excitation energy ($E_2^+$), deformation parameter ($\beta_2$ and $\beta_4$), and the Woods-Saxon (WS) parameters ($V_0$, $r_0$, $\&$ $a_0$)  of the target nuclei taken from Ref. \cite{rama01}}.
\renewcommand{\tabcolsep}{0.1cm}
\renewcommand{\arraystretch}{1.2}
\begin{tabular}{ccccccc}
\hline \hline 
System& $E_2^+$ & $\beta_2$ & $\beta_4$&$V_0$ & $r_0$ & $a_0$ \\
& (MeV) & & &(MeV) & (fm) & (fm)  \\
\hline
$^{18}$O+$^{74}$Ge  & 0.59 & 0.213 & -0.021 &56.46& 1.17 & 0.60\\  
$^{18}$O+$^{148}$Nd  & 0.30 & 0.201 & 0.078 &61.89& 1.16 & 0.60 \\
$^{18}$O+$^{182}$W & 0.10 & 0.265 & -0.075 &98.76& 1.15 & 0.73 \\
$^{18}$O+$^{186}$W & 0.12 & 0.226 & -0.095 &63.60& 1.18 & 0.66\\

\hline \hline 
\end{tabular}
\label{table2}
\end{table}

\begin{figure*}[h]
\begin{center}
\includegraphics[width=130mm,height=75mm,scale=1.5]{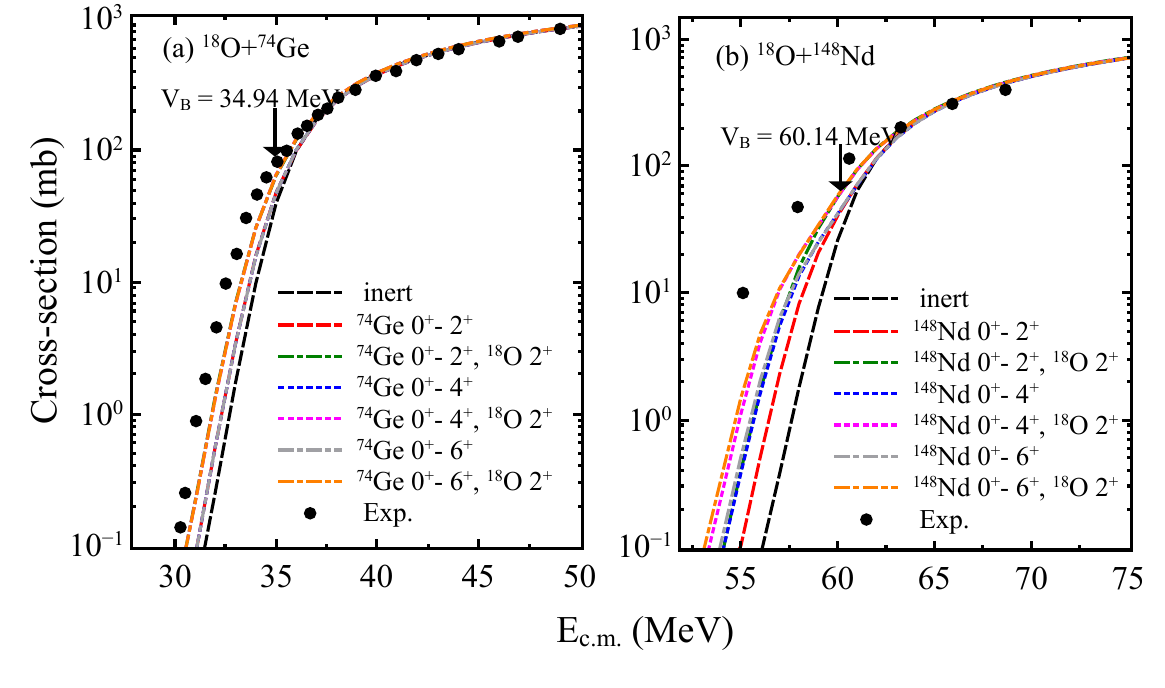}
\vspace{-0.3cm}
\caption{\label{fig1} (Color online) The fusion cross-section for $^{18}$O+$^{74}$Ge, and $^{18}$O+$^{148}${Nd} reactions. The dashed black lines in the graph depict the CC calculations conducted without any coupling mechanism, i.e., using 1D-BPM. The dashed red, dotted blue, and dotted dashed lines represent the $\sigma_{fus}$ corresponding to $2^+$, $4^+$, and $6^+$ states, respectively. The dotted dashed green, dotted magenta, and dotted dashed orange line represent the cross-section corresponding to $2^+$, $4^+$, and $6^+$ states with the inclusion of $^{18}$O as a projectile. $V_B$ is the barrier height represented by a downward arrow. The solid black circles represents the experimental data \cite{brod75,jia12}.}
\end{center}
\end{figure*}

This study seeks to explore the interrelation between the structure of a given system and its associated reactions. Specifically, we will investigate the impact of rotational or vibrational states of the target nuclei on the fusion cross-section ($\sigma_{fus}$) using the Coupled Channels (CC) approach implemented in the CCFULL code. For these calculations, we have employed the Woods-Saxon (WS) parameterization of the Aky$\ddot{u}$z-Winther parameter, with $V_0$, $r_0$, and $a_0$ selected to optimize the fit of $\sigma_{fus}$ at energies above the barrier. The potential parameters, along with the deformation and excitation energies corresponding to the $2^+$ state, are provided in Table \ref{table2}. The 1D-BPM calculations are depicted by the dashed black line in Fig. \ref{fig1}, exhibiting underestimation of experimental data, particularly at sub-barrier energies \cite{brod75, jish22}. Our findings indicate that the 1D-BPM calculations adequately account for $\sigma_{fus}$ in all considered reactions (i.e., $^{18}$O+$^{74}$Ge, $^{18}$O+$^{148}${Nd}, and $^{18}$O+$^{182}$W), except for the $^{18}$O + $^{186}$W reaction in the above-barrier energies region. The main aim of our study is not to address the fusion cross-section at energies above the Coulomb barrier but to reduce the hindrance at energies below the barrier. To address the observed fusion hindrance at sub-barrier energies, coupled-channel calculations are conducted with a deformed target and $^{18}$O as a vibrator. Initially, the $2^+$ individual rotational state of the target nuclei is considered, resulting in $\sigma_{fus}$ represented by the dotted red line. Our results illustrate the comparison between different excited states and the inert state (1D-BPM). Incorporating additional individual $4^+$ and $6^+$ states of the target nuclei further amplifies $\sigma_{fus}$, depicted by the dotted blue and dotted-dashed grey lines corresponding to the $2^{nd}$ and $3^{rd}$ excitation channels, respectively.
To assess the impact of excitation on the fusion cross-section, we consider the collective excitation state of the target nuclei in all analyzed reactions. In the case of reactions such as $^{18}$O + $^{74}$Ge and $^{18}$O + $^{148}$Nd, the fusion cross-section ($\sigma_{fus}$) undergoes a notable alteration upon incorporating the individual $2^+$ state of the target nuclei, as compared to the 1D-BPM model, as illustrated in Fig. \ref{fig1}(a) and \ref{fig1}(b). However, the inclusion of the $4^+$ state results in only a slight change in the fusion cross-section for the $^{18}$O + $^{148}$Nd reaction, while a negligible change is observed for the $^{18}$O + $^{74}$Ge reaction. This discrepancy is attributed to the relatively high values of $\beta_2$ and $\beta_4$ in reactions involving $^{148}$Nd target nuclei. Furthermore, considering the $6^+$ states leads to a modest increment in the fusion cross-section for the $^{18}$O + $^{148}$Nd reaction, while it remains negligible for the $^{18}$O + $^{74}$Ge reaction.
\begin{figure*}
\begin{center}
\includegraphics[width=130mm,height=75mm,scale=1.5]{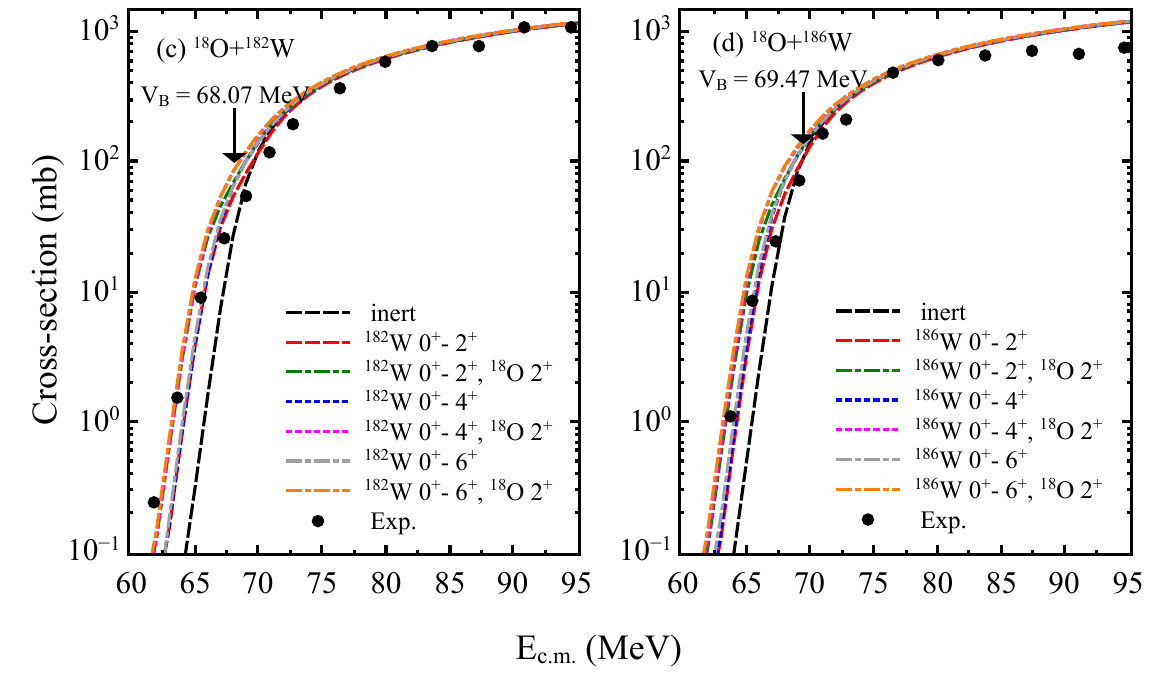}
\vspace{-0.3cm}
\caption{\label{fig2} (Color online) Same as Fig. \ref{fig1} but for $^{18}$O+$^{182}$W, $^{18}$O+$^{186}${W} reactions. The solid black circles represent the experimental data \cite{jish22}.}
\end{center}
\end{figure*}

On the other hand, for the $^{18}$O + $^{182}$W and $^{18}$O + $^{186}$W reactions, a notable change in the fusion cross-section is observed with the inclusion of the $2^+$ excited state of $^{18}$O as a projectile, as depicted in Fig. \ref{fig2}(a) and \ref{fig2}(b). However, due to the negative magnitude of $\beta_4$ in both reactions, no significant further change is observed with the collective excitation of $4^+$ and $6^+$ states, respectively. The variation in the magnitude of the hexadecapole deformation ($\beta_4$) \cite{jain23,jain24} explains this diverse behaviour, with $^{18}$O + $^{148}$Nd reaction featuring a positive magnitude and others exhibiting a negative magnitude (e.g., $^{18}$O + $^{74}$Ge, $^{18}$O + $^{182}$W and $^{18}$O + $^{186}$W). Despite the enhancement in $\sigma_{fus}$ compared to the 1D-BPM, the inclusion of the excited state still underestimates the experimental data, and the fusion hindrance persists. To mitigate this hindrance, the first vibrational state $2^+$ of $^{18}$O is also incorporated in the calculation. These results are corroborated by the step-by-step inclusion of coupling with other states in the CCFULL calculations, as illustrated in Fig. \ref{fig1} and  \ref{fig2}.\\ 
The expansion of CC calculations to include projectile excitations proves insufficient to replicate the data across all energy regions. The outcomes reveal that coupling the $2^+$ vibrational state of $^{18}$O with the $\beta_2$ and $\beta_4$ values of the target nuclei leads to an increase in $\sigma_{fus}$ compared to the entrance channel with rotational degrees of freedom of target nuclei alone. The $\sigma_{fus}$ with the involvement of the $2^+$ excited state of $^{18}$O is illustrated by the dotted-dashed green, dotted magenta, and dotted-dashed orange lines corresponding to the $2^+$, $4^+$, and $6^+$ states of the target, respectively, as depicted in Fig. \ref{fig1} and \ref{fig2}. Successful reproduction of experimental data is observed for the $^{18}$O + $^{182}$W reaction at energies above and below the Coulomb barrier, as illustrated in Fig. \ref{fig2}(a). However, it is important to note that including the vibrational state of the single phonon $2^+$ of $^{18}$O results in an overestimation of the experimental $\sigma_{fus}$ near and below the barrier, as shown in Fig. \ref{fig2}(b). Although it accurately reproduces the experimental $\sigma_{fus}$ at energies above the barrier for all considered reactions, it overestimates the results for the $^{18}$O + $^{186}$W reaction. However, fusion hindrance persists for the $^{18}$O + $^{74}$Ge, and $^{18}$O + $^{148}$Nd reactions, as depicted in Fig. \ref{fig1}(a, b). Overall, these results suggest that the addition of the $2^+$ vibrational state of $^{18}$O with the target nuclei enhances $\sigma_{fus}$ more compared to the excitation state of the target. However, the nucleon transfer channel which may significantly affect the below barrier fusion cross-sections are not considered in our present calculations.
\begin{figure*}
\begin{center}
\includegraphics[width=130mm,height=45mm,scale=1.5]{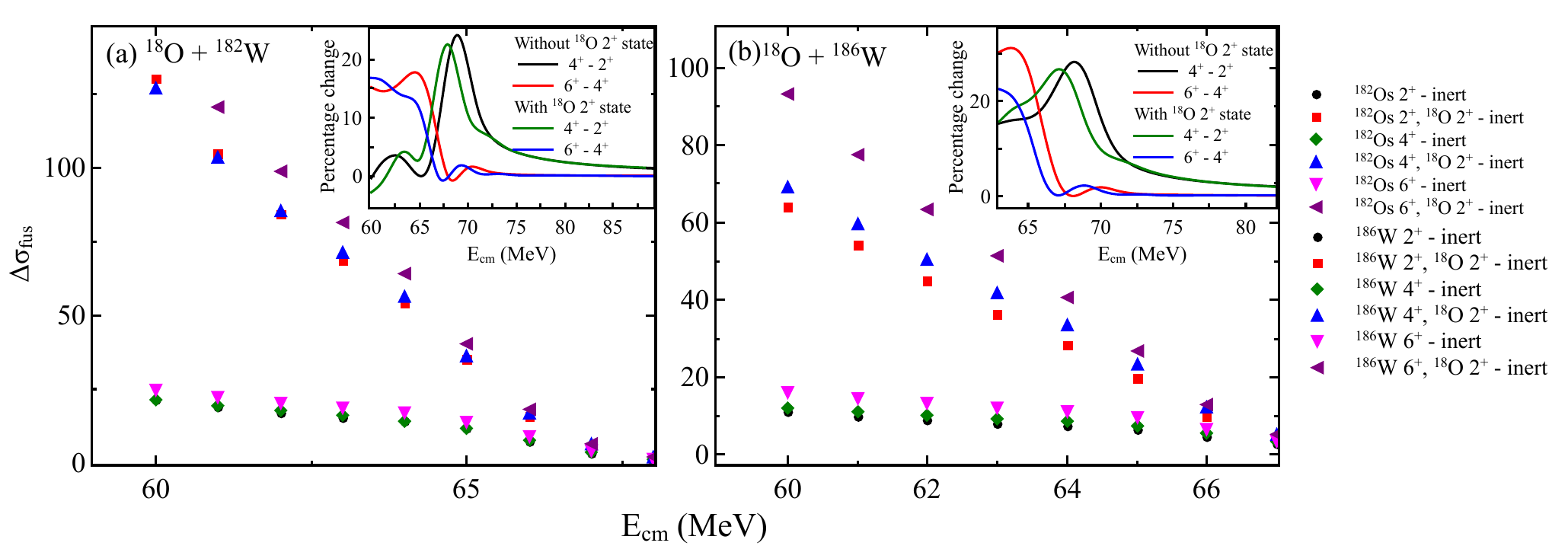}
\vspace{-0.3cm}
\caption{\label{fig3} (Color online) The relative and percentage change in the $\sigma_{fus}$ for (a) $^{18}$O+$^{182}${W}, and (b) $^{18}$O+$^{186}$W reactions.}
\end{center}
\end{figure*}

To determine the relative change $\Delta\sigma_{fus}$ in the $\sigma_{fus}$, one can use the following equation:
\begin{equation}
    \Delta\sigma_{fus} = \frac{(\sigma_{excited \ state} - \sigma_{inert})}{\sigma_{inert}}
\end{equation}
Our analysis involves the computation of $\Delta\sigma_{fus}$, representing the ratio of the difference between $\sigma_{excited \ state}$ and $\sigma_{inert}$ to $\sigma_{inert}$, incorporating multiple inelastic excitations and relative motion to accurately quantify $\Delta\sigma_{fus}$ in our CC calculations. Initially, we calculate $\Delta\sigma_{fus}$ for $^{18}$O-induced reactions, observing that all four systems exhibit a positive value of $\Delta\sigma_{fus}$ at below-barrier energies. A higher value of $\Delta\sigma_{fus}$ indicates an increase in the excitation state cross-section relative to the inert state. Furthermore, the absence of inelastic channels in the coupling necessitates the inclusion of additional coupling mechanisms to explain the sub-barrier $\sigma_{fus}$. In our study, we concentrate on the $^{18}$O + $^{182}$W, $^{18}$O + $^{186}$W reactions, considering that the relative change in $\sigma_{fus}$ corresponding to the $6^+$ state compared to the inert state is greater than that for other excited states like $2^+$ and $4^+$ states, as depicted in Fig. \ref{fig3}. In particular, we observe that the relative change $\Delta\sigma_{fus}$ is greatest at below-barrier energies and decreases as $E_{cm}$ increases, eventually becoming less than or approximately equal to zero at the highest barrier energies. It is imperative to incorporate additional excitation channels to elucidate the below-barrier $\sigma_{fus}$. In this study, we have introduced the $2^+$ state of $^{18}$O with the target nuclei, amplifying the relative change $\Delta\sigma_{fus}$ with respect to the inert state. In Fig. (\ref{fig3}a) and (\ref{fig3}b), the comparison between different excitation states relative to the ground state is presented. Specifically, there is an almost 98.71\% relative change at 62 MeV when the $2^+$ state of the projectile is coupled with the $6^+$ state of the target nuclei for the $^{18}$O + $^{182}$W reaction, and $\Delta\sigma_{fus}$ = 62.33\% for the $^{18}$O + $^{186}$W reaction. The percentage change between different excited states is also depicted in the sub-figures, with the solid black and green lines representing the percentage change in $\sigma_{fus}$ from the $2^{nd}$ excitation state to the $1^{st}$ excitation state without and with the inclusion of $^{18}$O having the $2^+$ excited state, respectively. In the case of the $^{18}$O + $^{182}$W reaction, the percentage change at 63 MeV is 3-4\% without and including the first excited state of $^{18}$O nuclei. Furthermore, at 83 MeV energy, the percentage change is almost negligible (around 1.5\%). Similarly, for the $^{18}$O + $^{186}$W reaction, the percentage change between the $2^{nd}$ and $1^{st}$ excited states is around 15-16\%, when considering the reaction without and including the first excited state of $^{18}$O nuclei at 63 MeV. However, as the center of mass energy increases, the percentage change between different excited states decreases. For instance, at 73 MeV, the difference is around 5-6\% without and with the inclusion of the $2^+$ vibrational state of $^{18}$O nuclei.

Similarly, the solid red and blue lines illustrate the relative changes of the $3^{rd}$ excitation state to the $2^{nd}$ excitation state without and with the inclusion of $^{18}$O having the $2^+$ excited state. In the case of the $^{18}$O + $^{182}$W reaction, the percentage differences at 63 MeV are 15.99\% and 13.98\% without and including the first excited state of $^{18}$O nuclei. Furthermore, at 83 MeV energy, the percentage differences are around 14-15\%. Similarly, for the $^{18}$O + $^{186}$W reaction, the percentage differences between the $3^{rd}$ and $2^{nd}$ excited states are 30.32\% and 22.50\%, respectively, when considering the reaction with and without the inclusion of the first excited state of $^{18}$O nuclei at 63 MeV. For instance, at 73 MeV, the differences are 0.46\% and 0.38\% without and with the inclusion of the $2^+$ vibrational state of $^{18}$O nuclei. However, as the center of mass energy increases, the percentage differences between different excited states decrease. The obtained results are consistent with those obtained for the other two reactions studied. The other reactions decrease in intensity similarly to these two reactions. It is important to note here that the nucleon transfer channels are not considered in the present calculations, which may play a significant role at energies below the Coulomb barrier.
\section{Summary and Conclusions}
\label{summary} 
In this study, we explored the influence of low-lying intrinsic degrees of freedom and their structural characteristics on the fusion cross-section for four reactions occurring at energies below the Coulomb barrier: $^{18}$O+$^{74}$Ge, $^{18}$O+$^{148}$Nd, $^{18}$O+$^{182}$W, and $^{18}$O+$^{186}$W. The fusion cross-section for each reaction at various energy levels is computed using the coupled channel code CCFULL. Our analysis primarily focuses on the energy levels of the involved nuclei, with special attention to their rotational and vibrational states. The results indicate a good agreement between theoretical calculations and experimental data for the nuclei  $^{74}$Ge, $^{148}$Nd, $^{182}$W, and $^{186}$W, particularly for the $2^+$ excited state. While minor deviations are observed for other excited states ($4^+$ and $6^+$), the overall agreement remains substantial. Moreover, our results highlight the significant impact of mixed hexadecapole deformation on the fusion cross-section. When $\beta_4$ possesses a positive magnitude, rotational levels beyond $6^+$ no longer contribute significantly, leading to a marked difference in the contribution of sequential channels. Conversely, for a negative value of $\beta_4$, rotational energy levels up to the $2^+$ state play a crucial role in influencing fusion characteristics. To further validate our analysis, we compare our computed results with experimental fusion excitation data. Additionally, we calculate the relative change ($\Delta\sigma_{fus}$) between ground and excited states with and without coupling terms. This comprehensive investigation provides valuable insights into the intricate interplay between nuclear structure and dynamics in heavy-ion fusion reactions, contributing to a deeper understanding of these fundamental processes.\\ \\

\section*{Acknowledgments}
\noindent
This work is partially supported by the Science and Engineering Research Board (SERB) File No. CRG/2021/001229, and Ramanujan Fellowship File No. RJF/2022/000140.

\end{document}